# The Relativistic Transactional Interpretation and The Quantum Direct-Action Theory

R. E. Kastner[*]
Foundations of Physics Group, University of Maryland
29 December 2020

ABSTRACT. This paper presents key aspects of the quantum relativistic direct-action theory that underlies the Relativistic Transactional Interpretation. It notes some crucial ways in which traditional interpretations of the direct-action theory have impeded progress in developing its quantum counterpart. Specifically, (1) the so-called 'light tight box' condition is re-examined and it is shown that the quantum version of this condition is much less restrictive than has long been assumed; and (2) the notion of a 'real photon' is disambiguated and revised to take into account that real (on-shell) photons are indeed both emitted and absorbed and therefore have finite lifetimes. Also discussed is the manner in which real, physical non-unitarity naturally arises in the quantum direct-action theory of fields, such that the measurement transition can be clearly defined from within the theory, without reference to external observers and without any need to modify quantum theory itself. It is shown that field quantization arises from the non-unitary interaction.

1 Introduction.

This paper concerns the relativistic extension and elaboration of the Transactional Interpretation (TI) of quantum mechanics, first proposed by Cramer (1986). TI is based on the Wheeler-Feynman "absorber" or "direct-action" theory of fields ((Wheeler and Feynman, 1945, 1949). In the direct-action theory (DAT), the elementary interaction between charges is time-symmetric in character, and absorbers respond to emitters with a time-symmetric field that is exactly out of phase with the emitted field. For an overview of key terminology and concepts of TI, see Kastner (2016a). Briefly, in TI, the usual quantum state or "ket" is referred to as an "offer wave" (OW), while the response of an absorber, an advanced field represented by a "bra", is called a "confirmation wave."

The Relativistic Transactional Interpretation (RTI), as developed by the present author (Kastner 2018 and references therein) finds its natural theoretical underpinning in the quantum relativistic direct-action theory of Paul Davies (1971, 1972). Like the classical Wheeler-Feynman theory, the Davies theory describes interactions among systems in terms of a direct connection between currents (field sources) rather than by way of a mediating field with independent degrees of freedom. One of the original motivations for such an "action at a distance" theory was to eliminate divergences, stemming from self-action of the field, from the standard theory. However, it was later realized that some form of self-action was needed in order to account for such quantum phenomena as the Lamb shift. The Davies theory does allow for self-action in that a current can be regarded as acting on itself in the case of indistinguishable currents (see, e.g., Davies (1971), p. 841, figure 2).

---

[*] rkastner@umd.edu

Nevertheless, despite RTI's natural affinity for the direct-action theory of fields, it must be emphasized that RTI does *not* involve an ontological elimination of the field. The field concept remains as a measure of the direct interaction between charges, quantifiable in terms of a potential. The basic field is non-quantized, but quantization emerges under suitable conditions. We return to this issue in Section 5.

Thus, RTI is based not on elimination of fields, but rather on the time-symmetric, transactional character of energy propagation by way of those fields, and the feature that offer and confirmation waves capable of resulting in empirically detectable transfers of physical quantities only occur in couplings between field currents. However, the fields themselves are considered as pre-spacetime objects. That is, they exist; but not as spacetime entities. Rather, spacetime entities are restricted to actualized, detectable conserved currents: real-valued energy/momentum transfers. At first glance, this ontology may seem strange; however, when one recalls that such standard objects of quantum field theory as the vacuum state |0> have no spacetime arguments and are maximally non-local,[1] it seems reasonable to suppose that such objects exist, but not in spacetime (in the sense that they cannot be associated with any region in spacetime).

Shimony (2009) has similarly suggested that spacetime can be considered as a domain of actuality emergent from a quantum level of possibilities:

> There may indeed be "peaceful coexistence" between Quantum nonlocality and Relativistic locality, but it may have less to do with signaling than with the ontology of the quantum state. Heisenberg's view of the mode of reality of the quantum state was. . . that it is *potentiality* as contrasted with *actuality*. This distinction is successful in making a number of features of quantum mechanics intuitively plausible — indefiniteness of properties, complementarity, indeterminacy of measurement outcomes, and objective probability. But now something can be added, at least as a conjecture: that the domain governed by relativistic locality is the domain of actuality, while potentialities have *careers* in space-time (if that word is appropriate) which modify and even violate the restrictions that space-time structure imposes upon actual events. (2009, section 7, item 2)

Shimony goes on to note the challenges in providing an account of the emergence of actuality from potentiality, which amounts to "collapse" or quantum state reduction. RTI suggests that transactions are the vehicle for this process;[2] and that aspects of it must involve

---

[1] This is demonstrated by the Reeh–Schlieder Theorem; cf. Redhead (1995).
[2] Recall that even if no specific "mechanism" is provided for the actualization of a transaction, TI provides a solution to the measurement problem in that it ends the usual infinite regress by taking into account absorption, which physically determines the measurement basis. A measurement is completed when absorption occurs, and the conditions for that can be precisely specified, as shown herein. Moreover, as suggested above, it is likely

processes and entities transcending the spacetime construct. It thus differs from one aspect of Shimony's formulation, in that potentialities are taken as existing outside the spacetime manifold; so they have "careers" in an extra-spatiotemporal quantum substratum, rather than in spacetime—and that is why they evade some of the restrictions of relativity.

A further comment is in order regarding the proposal that spacetime is emergent rather than fundamental. In the introductory chapter to their classic *Quantum Electrodynamics*, Beretstetskii, Lifschitz, and Petaevskii make the following observation concerning QED interactions:

> For photons, the ultra-relativistic case always applies, and the expression $[\Delta q \sim \hbar / p]$, where Δq is the uncertainty in position, is therefore valid. This means that the coordinates of a photon are meaningful only in cases where the characteristic dimension of the problem is large in comparison with the wavelength. This is just the "classical" limit, corresponding to geometric optics, in which the radiation can be said to be propagated along definite paths or rays. In the quantum case, however, where the wavelength cannot be regarded as small, the concept of coordinates of the photon has no meaning. . .
>
> The foregoing discussion suggests that the theory will not consider the time dependence of particle interaction processes. It will show that in these processes there are no characteristics precisely definable (even within the usual limitations of quantum mechanics); *the description of such a process as occurring in the course of time is therefore just as unreal as the classical paths are in non-relativistic quantum mechanics*. The only observable quantities are the properties (momenta, polarization) of free particles: the initial particles which come into interaction, and the final properties which result from the process. [The authors then reference L. D. Landau and R. E. Peierls, 1930.[3]](Emphasis added.) (1971, p. 3)

The italicized sentence asserts that the virtual particle interactions described by QED (and, by extension, by other interacting field theories) cannot consistently be considered as taking place in spacetime. Yet they do take place *somewhere*; the computational procedures deal with entities implicitly taken as ontologically real. This "somewhere" is just the quantum substratum alluded to above.

## 2. Background

In this section, we will first review the basic classical absorber or 'direct action' theory and a semi-classical quantum version due to Davies (1971, 1972). It should be noted that Davies' treatment, while an advance in the quantum direction from the original classical Wheeler-Feynman theory, remained semi-classical insofar as it tacitly identified radiation with continuous fields. It also assumed that a real photon could be unilaterally emitted, which, as we shall see, is not the case at the quantum level. Thus, ambiguity remained in that account

---

misguided to demand a causal, mechanistic account of collapse, since as Shimony suggests, one is dealing with a domain that transcends the causal spacetime realm.

[3] The Landau and Peierls paper has been reprinted in Wheeler and Zurek (1983).

regarding the distinction between real and virtual photons, as well as the nature of the relevant absorber boundary condition, or so-called 'light tight box' condition, which has led to some confusion. However, it is a useful starting point for the present work, which revises certain features pertaining to the quantization of the radiated field. The revised account makes clear the fully quantum nature of the appropriate boundary condition, which is really a particular sort of emitter/absorber interaction rather than any specific configuration of absorbers as is implied by the usual term "light-tight box."

We first review standard classical electromagnetic theory. The standard way of representing the field *A* acting on an accelerating charge *i* due to other charges *j* is as the sum of the retarded fields due to *j* and a 'free field':

$$A = \sum_{j \neq i} A^{ret}_{(j)} + \frac{1}{2}\left(A^{ret}_{(i)} - A^{adv}_{(i)}\right) \qquad (2)$$

In the classical expression (2), self-action is omitted to avoid infinities (which are dealt with in quantum field theory by renormalization). $A^{ret}_{(j)}$ is the retarded solution to the inhomogeneous equation, i.e., the field equation with a source, while the second term pertaining only to *i* is a solution to the homogeneous field equation (source-free). The latter quantity, the 'radiation term,' is originally due to Dirac and is necessary in order to account for the loss of energy by a radiating charge if it is assumed that all sourced fields are retarded only. Wheeler and Feynman (1945) critically remark in this regard:

> "The physical origin of Dirac's radiation field is nevertheless not clear. (a) This field is defined for times before as well as after the moment of acceleration of the particle. (b) The field has no singularity at the position of the particle and by Maxwell's equations must, therefore, be attributed either to sources other than the charge itself or to radiation coming in from an infinite distance." (p. 159)

These authors' concern about the source of Dirac's radiation field is resolved in the direct-action theory (henceforth DAT). The classical direct-action or 'absorber' theory proposed that the total field *A*<sup>(DA)</sup> acting on *i* is given by:

$$A^{(DA)} = \sum_{j \neq i} \frac{1}{2}\left(A^{ret}_{(j)} + A^{adv}_{(j)}\right) \qquad (3)$$

i.e., it is given by the sum of the time-symmetric fields generated by all charges *except i*. Absorbing charges respond to the emitted field with their own time-symmetric field, contributing to the sum in (3). Wheeler and Feynman noted that (2) and (3) are equivalent provided that their difference is zero, i.e.:

$$\sum_{\forall j} \frac{1}{2}\left(A_{(j)}^{ret} - A_{(j)}^{adv}\right) = 0 \tag{4}$$

Under the condition (4), the responses of absorbing charges to the time-symmetric field of the emitting charge yields an effective 'free field' applying only to the emitting charge; i.e. the second term of (2). It's important to note that this term attributes a solution to the homogeneous equation to a particular charge that is (of course) not its source, as observed by WF above. In the DAT, the 'free field' is actually sourced by other charges (responding absorbers) and only *appears* to have the form of a free field from the standpoint of the accelerating charge whose index it bears.

The condition (4) is historically termed the 'light tight box' condition (LTB) in the classical theory. It is commonly interpreted as the constraint that 'all radiation is absorbed,' but this characterization is misleading even at the classical level, and requires explicit reformulation at the quantum level. For one thing, it conflates the static, time-symmetric Coulomb field with a dynamic radiation field.[4] In addition, the mathematical content of (4) says only that the net radiation field is zero. This can just as legitimately be interpreted to mean that *there is no true free (unsourced) radiation field*. While selective cancellation of fields does occur among charges to produce the effective local radiation field, the absence of an unsourced radiation field is the primary physical content of the "LTB" condition for the quantum form of the DAT, as we will see in the next section.

Other weaknesses in the original classical DAT have been discussed by Gründler (2015), who notes that field cancellation via explicit evaluation of the interactions between the emitter and the other charges depends on imposing an arguably unjustified asymmetrical condition: an effective index of refraction applying only to absorber responses. He argues that the equivalence between the classical DAT and standard classical electrodynamics for individual charges amounts only to a formal one based on (3) and (4).

In any case, the ambiguity inherent in the classical treatment, and the practice of interpreting (4) as being about some specific distribution of charges, has led to some confusion regarding the nature of the relevant condition -- the analog of (4) -- pertaining to the quantum case. In what follows, we clarify the nature of the quantum version of the direct action theory, or "QDAT" for short, and define the applicable boundary conditions.

## 3. The quantum direct-action theory: basics

In this section, we will discuss the DAT in terms of Green's functions or 'propagators' (solutions to the field equation for a point source, and related source-free forms), since that is the natural way to formulate the QDAT. It should be noted that, in contrast to the field *A(x)* with a single argument, propagators are functions of two arguments, and always relate two specific coordinate points. In standard quantum field theory, propagators are correlation functions for pairs of field coordinates.[5] In the QDAT, propagator arguments are parameters of the source currents (charges).

---

[4] Actually, the classical DAT appears to assume that even the time-symmetric fields are present only in the case of an accelerating charge, which neglects the static Coulomb interaction.

The corresponding quantities are:

$D_{ret}(x-y)$: retarded solution to the inhomogeneous equation

$D_{adv}(x-y)$: advanced solution to the inhomogeneous equation

$\bar{D}(x-y) = \frac{1}{2}(D_{ret} + D_{adv})$: time-symmetric solution to the inhomogeneous equation

$D(x-y) = (D_{ret} - D_{adv})$: odd solution to the homogeneous equation

In terms of these, we can see that the following identity holds:

$$D_{ret} = \bar{D} + \frac{1}{2} D \qquad (5)$$

This describes the elementary field of a single charge in the DAT, taking into account the "response of the absorber" corresponding to the second term. It differs from (2) in that it does not exclude the charge from the effects of the field. As noted by Wheeler and Feynman (1945), the first term is singular, which is why they sought to prohibit a charge from interacting with its own field. At the quantum level, in view of indistinguishability, one cannot impose such a restriction, since in general one cannot define which charge is the source of the time-symmetric field.[6] So the self-action involving the time-symmetric field must be retained at the quantum level (however, we will see later that this self-action does not involve any exchange of real energy and consists only of self-force). This expression shows how a net retarded field arises due to the combination of absorber response (an effective "free field" acting on the emitting charge) with the basic time-symmetric field of the emitting charge. We now investigate the analogous situation in the QDAT.

First, it is important to note that the propagators defined above make no distinction between positive and negative frequencies, since the classical theory makes no connection between frequency and energy. However, the quantum theory of fields must explicitly deal with the existence of positive and negative frequencies. Thus, in the QDAT, each of the quantities above must be understood as comprising positive- and negative-frequency components. Since there are many different conventions for defining these quantities, we write the components here explicitly in terms of vacuum expectation values or 'cut propagators' $\Delta^{\pm}$. In these terms,

$$iD(x-y) \equiv i\Delta(x-y) = \langle 0|[A(x), A(y)]|0\rangle$$
$$= (\langle 0|A(x)A(y)|0\rangle - \langle 0|A(y)A(x)|0\rangle) \equiv (\Delta^+ - \Delta^-) \quad (6)$$

where $A(x)$ is the usual quantum electromagnetic field, and under Davies' convention for the components, we define

$$D(x-y) = D^+ + D^- = (-i\Delta^+) + (i\Delta^-) \quad (7)$$

---

[5] As suggested by Auyang (1995), these coordinates are best understood as parameters of the field, rather than 'locations in spacetime.' The same understanding can be applied to the non-quantized field of the QDAT, in which field sources are the referent for the parameters.

[6] Charges become distinguishable only in situations in which energy-momentum conservation is satisfied and non-unitarity can occur. This issue is elaborated in Section 5.

Note in particular, for later purposes, that $D^-$ is defined with the opposite sign of the negative-frequency cut propagator $\Delta^-$:

$$iD^-(x-y) \equiv -\Delta^-(x-y) = -\langle 0|A(y)A(x)|0\rangle \quad (8)$$

We also need the even solution to the homogeneous equation, $D_1$ (cf. Bjorken and Drell, 1965, Appendix C):

$$D_1(x-y) = i(D^+(x-y) - D^-(x-y)) = \Delta^+(x-y) + \Delta^-(x-y) \quad (9)$$

Note that each of the positive- and negative-frequency components of these fields independently reflects the same relationship of retarded and advanced solutions as the total field; e.g., $D^+(x-y) = (D^+_{ret} - D^+_{adv})$.

Feynman's innovation was to interpret negative frequencies as antiparticles; specifically, as "particles with negative energies propagating into the past." This is equivalent to antiparticles with positive energy propagating into the future, where antiparticles have the opposite charge (cf. Kastner 2016b). To that end, he defined a propagator that does just that, i.e. assigns the retarded propagator only to positive frequencies and the advanced propagator only to negative frequencies. The result is the "Feynman propagator," $D_F$:

$$D_F = D^+_{ret} + D^-_{adv} \quad (10)$$

This satisfies an identity analogous to (5):

$$D_F = \bar{D} - \frac{i}{2} D_1 \quad (11)$$

To see (11) explicitly, we write the quantities in terms of their positive- and negative frequency components, using (9) for $D_1$ :

$$\bar{D} - \frac{i}{2} D_1 = \frac{1}{2}\left[\left(D^+_{ret} + D^+_{adv}\right) + \left(D^-_{ret} + D^-_{adv}\right)\right] + \frac{1}{2}\left[\left(D^+_{ret} - D^+_{adv}\right) - \left(D^-_{ret} - D^-_{adv}\right)\right]$$
$$= \left(D^+_{ret} + D^-_{adv}\right) = D_F \quad (12)$$

## 4 'Light tight box' condition modified at quantum level

As observed by Davies (1971), a basic quantum version of the direct-action theory (QDAT) has actually been around since Feynman (1950). Feynman showed that for the case when the number *n* of external (commonly termed "real") photon states is zero, the standard quantum action *J* for the interaction of the quantized field $\hat{A}$ with a current *j* can be replaced by a direct current-to-current interaction, as follows:

$$J(n=0) = \sum_i \int j^\mu_{(i)}(x)\hat{A}_\mu(x) \, d^4x =$$

$$\sum_i \sum_j \frac{1}{2} \int j^\mu_{(i)}(x) D_F(x-y) j_{\mu(j)}(y) \, d^4x \, d^4y \quad (13a,b)$$

where $D_F$ is the Feynman propagator as defined in (10) and (11). Davies notes that the same result is proved by way of the S-matrix in Akhiezer and Berestetskii (1965), p. 302 (henceforth 'AB'). So it is important to note that (13) *is a theorem,* and holds even if one has started from the usual assumption that there exists an independent quantized electromagnetic field $\hat{A}$.

Now, the entire content of the so-called "light tight box condition" (LTB) for the quantum version of the direct action theory (QDAT) is contained in the condition for the equivalence of the two expressions (a) and (b) in (13). But the LTB condition has traditionally been deeply mired in ambiguity about what sort of entity counts as a "real photon," and about what physical situations give rise to real photons. It has additionally been hampered by a semi-classical notion of "absorption of radiation." However, it is straightforward from the mathematics that what is actually required for the equivalence of the two expressions in (13) is simply the non-existence of an independent quantized electromagnetic operator field $\hat{A}$ -- i.e., vanishing of the usual postulated system of oscillators of standard quantum field theory! We can see that explicitly by way of the proof of AB, who obtain an expression for the scattering matrix *S* in the general case, with no restriction. That expression is:

$$S = \exp\left(-\frac{i}{2}\int j^\mu(x) D_F(x-y) j_\mu(y) \, d^4x \, d^4y\right) \times \exp\left(i\int j_\mu(x)\hat{A}^\mu(x) \, d^4x\right) \quad (14)$$

where the usual chronological ordering of quantum field operators is understood, and $\hat{A}^\mu$ is the usual quantized electromagnetic field. AB then say: "In processes in which no photons participate, the last factor is equal to unity, and the scattering matrix assumes the form [first factor only, as in eqn. (13b)]." But again, this drags in the ill-defined notion of "participation of photons," when what is really done to obtain the final result is to simply *set the independent quantized electromagnetic field* $\hat{A}^\mu$ *to zero*. The crucial point, then, is the following: essentially all there is to the so-called "light tight box" condition for the QDAT expressed in terms of the Feynman propagator $D_F$ is *Wheeler and Feynman's original proposal to eliminate the electromagnetic field as an independent mechanical system.* Note that this corresponds to the condition (4) as interpreted in the previous section; i.e., that there simply are no genuinely unsourced "free fields." Rather, any effective field of the form $D$ (or $D_1$ for the QDAT) is obtained through a specific kind of interaction between sources, i.e., between emitters and absorbers.[7]

In the next section we examine the QDAT in more detail, resolving some ambiguities about the distinction between real and virtual photons and discussing the relevance of the distinction for the quantum form of the LTB condition. We'll see that the only additional condition for equivalence of the QDAT with the standard theory amounts to the quantum completeness condition (and an appropriate phasing of the fields of the emitter and absorbers), which assures recovery of the Feynman propagator $D_F$.

## 5  Relativistic generalization of absorber response

The Feynman propagator $D_F$ is the quantum analog of (2); it reflects a "causal" field directed from smaller to greater temporal values for the case of positive frequencies and the opposite--from greater to smaller temporal values--for negative frequencies, with an effective "free field" for radiative processes. $D_F$ arises due to the quantum relativistic analog of absorber response, which differs from the classical theory in two crucial respects. One is the need to take into account negative frequencies not present in the classical case, which requires separate phasing of the positive- and negative-frequency field components and leads to $D_1$ rather than D the free field, as discussed above. The other is the mutual, or dynamically symmetric, emitter/absorber interaction giving rise to the "free field" $D_1$. To clarify the second point: at the relativistic level (which is the level at which Nature really operates), emitters and absorbers participate *together* in the generation of offers and confirmations. Offers (OW) are not emitted unilaterally and *then* responded to; instead, *both* OW and CW are generated in a more symmetrical, mutual interaction that is non-unitary.[8] Importantly, this non-unitary interaction—the generation of offers and confirmations giving rise to a real, on-shell photon described by $D_1$—must be carefully distinguished from the basic time-symmetric field connection $\bar{D}$, which is unitary. The latter corresponds *only* to an off-shell, virtual photon, i.e., to the Coulomb field (zeroth component of the electromagnetic potential).

---

[7] Several authors have noted that one need not view the 'zero-point energy' as evidence for an independently existing field, since exactly the same effects attributed to zero-point energy arise in the QDAT. See, for example, Bennett (1987), Jaynes (1977).

[8] As gauge bosons, real (on-shell) photons are only created through the non-unitary process described herein. Fermionic sources can be on-shell in the absence of a non-unitary process; thus, one can have a real electron without an "electron CW." This important distinction between bosons and their fermionic sources is elaborated in Kastner (2019a).

For clarity in the discussion regarding which process is under consideration, let us use the term "U-interaction" to denote the unitary, Coulomb, virtual photon interaction described by $\bar{D}$, and the term "NU-interaction" to denote the non-unitary, radiative, real photon interaction described by $D_1$. The latter is the analog of "absorber response" at the non-relativistic level. The former, basic U-interaction obtains in situations that do not satisfy energy conservation; e.g., between two free electrons that would not be able to transfer real energy between them. The U-interaction conveys only force, not energy. In contrast, a NU-interaction corresponds to radiative processes only; i.e., to transversely polarized real photons described by Fock states (more precisely, projection operators; this point is elaborated below). The latter type of process occurs with a well-defined probability—basically a decay rate. It occurs only when energy-momentum (and angular momentum) conservation is satisfied. Under these conditions, participating charges attain distinguishability, in that one is clearly losing conserved quantities and others are (possibly) gaining them. Here, we must say "possibly" because many absorbers are responding with CW, but in the case of a single photon, only one absorber can actually gain the conserved quantities transferred. (This issue, involving probabilistic behavior, is elaborated further below).

Thus, in the QDAT, whenever there is only a static Coulomb field, it means that non-unitarity has *not* occurred; this is a U-interaction. The virtual photons that mediate the Coulomb field are not Fock states and thus are not described by offer or confirmation waves. For virtual photons, there is no fact of the matter about which current emitted and which current absorbed, since there is no OW or CW at this level. It is a force-only, symmetrical interaction and is not radiative. A useful mnemonic for this distinction is "virtual photons convey only force, while real photons convey energy." The fact that the unitary, time-symmetric connection conveys only force explains why any divergences associated with the self-interaction do not involve real energy; they are force-divergences only.[9]

In introducing this concept of the generation of a real photon—the NU-interaction--we come to an important previously 'missing link' in extending the transactional picture to the relativistic level. This observation addresses and resolves a common criticism that emitters and absorbers are "primitive" and that absorber response is just a placeholder for the "external observer" in the measurement problem. On the contrary, the behavior of emitters and absorbers that trigger the non-unitary measurement transition is not "external" to the theory at all. It is fully accounted for and quantified within the relativistic QDAT in terms of the coupling amplitude or charge *e*. This issue is discussed in detail in Kastner (2018) and in Kastner and Cramer (2018) and is further elaborated below. For now, we note that the charge *e* is the basic amplitude for a photon to be emitted or absorbed, as previously observed by Feynman (1985). In the context of RTI and the QDAT, it is the amplitude for either an OW or CW to be generated. Since, as described above, one needs *both* the OW and CW to create a real photon (corresponding to a Fock state projection operator) in the QDAT, there are two factors of the charge *e*; hence the basic probability of real photon generation—the NU-interaction— is the fine structure constant $\alpha = e^2$. The foregoing highlights the crucial physical role of the fine structure constant in governing radiative processes. As noted above, when the NU-interaction does not occur, one still has the basic time-symmetric connection $\bar{D}$ corresponding to a virtual (off-shell) photon mediating to the static Coulomb field. Thus, a field can certainly be generated as the basic connection $\bar{D}$ between currents, but with no radiation (no real photon and no real energy). The crucial point: *field generation in the QDAT is not necessarily radiated energy*. Radiation is emitted *only* if the NU-interaction occurs, and it need not and often does not occur.

---

[9] This arguably also provides an ultimate ontological basis for Newton's Third Law.

Another important distinction between the classical DAT and the QDAT is that the relevant quantity for describing the interaction is the scattering matrix $S = Pe^{-iJ}$ (where *J* is the action and *P* a time-ordering operator). *S* defines probability amplitudes for transitions between initial and final states. This probabilistic behavior does not exist at all in the classical DAT, but is a crucial aspect of the QDAT. Its importance in differentiating the quantum situation from the original WF theory cannot be overstated. Failure to appreciate the entry of quantum probabilities into the field behavior leads to great confusion regarding what is meant by terms like "emission" and "radiation." In particular, in the quantum case one must distinguish between (i) the generation of a field, which could be just the static Coulomb field mediated by virtual photons through the bound field $\bar{D}$ (the U-interaction) and (ii) the actual emission or radiation of a real (transversely polarized) photon, which occurs only for the NU-interaction giving rise to the "free field" $D_1$. In the classical case, there is no distinction between generating a field and radiating, since it is assumed that absorber response *always* occurs, and that all generated fields are radiative in nature—i.e., that they convey electromagnetic energy corresponding to the intensity of the field. However, this is *not* the case at the quantum level, since (as noted above) the $\bar{D}$ field can exist as a basic connection between currents without any corresponding radiation or energy transfer.

The need for a probabilistic description arises because in the quantum case, one must take into account that the field itself is not equivalent to a "photon" in that a photon is discrete while the field is continuous (at least with respect to the parameter *x*). As an illustration, suppose we are dealing with a field state corresponding to one photon. Such a field in general propagates between an emitter and many absorbers; many absorbers can respond, even though there is only one photon "in the field." While the responses contribute to the creation of the real photon through the NU-interaction, the photon itself cannot go to all the responding absorbers; only one can actually receive it. This is where the probabilistic behavior, described by $S = Pe^{-iJ}$, enters. We make this issue more quantitative in what follows.

Looking at the Fourier components, one again sees that the Feynman propagator is complex, with both real and imaginary parts:

$$D_F(x) = \frac{1}{(2\pi)^4} \int \left( \frac{PP}{k^2} - i\pi\delta(k^2) \right) e^{ikx} dk = \bar{D}(x) - \frac{i}{2} D_1(x) \qquad (15)$$

('PP' stands for the principal part.) The complexity of $D_F$ implies intrinsic non-unitarity, a point whose implications we will consider in §6. The real part $\bar{D}$ is the time-symmetric propagator, while the imaginary part $D_1$ is the even "free field" or solution to the homogeneous equation as defined above.[10]

As Davies notes, "The $\bar{D}$ part (bound field) leads to the real principal part term which describes virtual photons ($k^2 \neq 0$), whilst the imaginary part $D_1$ (free field) describes photons with $k^2 = 0$, that is, real photons, through the delta function term." (Davies 1972, p. 1027). The $D_1$ term is the quantum analog of the free field in eqns. (2) and (5). In the classical DAT, the "free field" is assumed to be present for all accelerated particles due to the "response of the universe" or "absorber response." In order to understand the circumstances and physical meaning of the $D_1$ interaction for the QDAT, we must clearly define the quantum analog of acceleration and distinguish that from the static case, in which only the Coulomb (non-radiative) interaction $\bar{D}$ is present. The quantum analog of acceleration is a state transition, such as from a higher to a lower atomic energy state, accompanied by radiation. In contrast, for the static case, there is no radiation, so there is no effective free field-- no "absorber response." Thus, in the QDAT, the presence or absence of "absorber response" -- really the mutual NU-interaction, as discussed above-- is what dictates whether there will be a $D_1$ component and hence a quantum form of acceleration accompanied by radiation (i.e., the exchange of transversely polarized, real photons). Without the NU-interaction, one still has the basic time-symmetric U-interaction; i.e., one has virtual photon exchange but not real photon exchange. As noted above, and as discussed in Kastner (2018) and Kastner and Cramer (2018), the basic probability of the occurrence of the NU-interaction and real photon transfer via the $D_1$ component is the fine structure constant.

In contrast, traditional quantum field theory (QFT) uses the entire $D_F$ universally. In view of the distinct physical significance of the real and imaginary part of the Feynman propagator as noted above, which holds regardless of the specific model considered, a shortcoming of traditional QFT is that no physical distinction can be made in that theory between radiative and non-radiative processes at the level of the propagator.[11] Indeed, in standard QFT the term "virtual photon" is routinely taken as synonymous with an internal line in a Feynman diagram. This is inadequate and misleading, as it is only a contextual criterion (depending on "how far out we look") and thus does not describe the photon itself. While Davies' definition quoted above—virtual photon is off the mass shell and corresponds to the time-symmetric propagator, while real photon is on the mass shell and corresponds to the free field term— is the correct account of the physical distinction between real and virtual photons, his treatment of the real/virtual distinction in both Davies (1971) and (1972) falters into an ambiguous one alternating between (a) the standard, inadequate QFT characterization of the real vs. virtual distinction as a merely contextual one, i.e. as an "internal" vs "external" photon dependent on our zoom level and (b) the mistaken assumption that a real photon must have an infinite lifetime and therefore can only be truly external.[12] In particular, he appears to apply the uncertainty principle to the lifetime of real photons. However, a real photon obeys energy conservation, and its lifetime is therefore not limited in that way.[13] The fact that real photons are *both* emitted and absorbed, and therefore can be considered a form of "internal line," is key in understanding the relevant quantum analog of the LTB condition. Indeed, all real photons are "internal" in the QDAT, since real photons can only be created through the participation of both emitters *and* absorbers.

---
[10] Here, we are using the sign conventions in Bjorken & Drell (1965), Appendix C.
[11] This issue is also the root cause of inconsistency problems in standard QFT, as revealed by Haag's Theorem. For details on how the QDAT can resolve such problems, see Kastner (2015).

So, keeping in mind that it is indeed possible to have a "real but internal" photon, let us review another useful account given in Davies (1971) of the relevant LTB condition for the QDAT. Davies correctly notes that the fully quantum form of the LTB is simply the requirement that there are no transitions between *external* fermion/photon states $|\beta\rangle = |\psi, n\rangle$ for which the photon number $n \neq 0$. He writes this as:

$$\sum_{\beta'} |\langle \beta' | S | \alpha \rangle|^2 = 0 \qquad (16)$$

where $|\alpha\rangle$ are states with n=0 and $|\beta'\rangle$ are states with $n \neq 0$. This is in keeping with the theorem (13) and the discussion of (14) above. But of course, the transition probability for each value of $\beta'$ is a non-negative quantity, so each term must vanish separately:

$$|\langle \beta' | S | \alpha \rangle|^2 = 0, \forall \beta' \qquad (17)$$

Also, note that by symmetry the restriction on external photon states $n \neq 0$ holds for both initial states and final states. That is, one must exclude transitions *from* states $|\alpha'\rangle$ as well as transition *to* states $|\beta'\rangle$. *Thus, the QDAT describes a world in which there simply are no truly external photons.* This, of course, simply corresponds to setting the independent quantized

---

[12] Davies notes that real photons are massless, but this just means they do not decay into other quanta. It does not preclude them from being emitted and absorbed. However, Davies does correctly criticize Feynman's purely contextual account of the 'real vs virtual' distinction by noting that a true virtual photon has no well-defined direction of energy transfer and is described by the time-symmetric component of $D_F$ (i.e. the real part $\bar{D}$) only (Davies 1972, p. 1028).

[13] Even if one disputed this, it is well known that emission and absorption require a finite spread in the energy level. So one cannot argue that a photon has an infinite lifetime because it has a definite energy; it is possible for a real photon to have a spread in energy. Davies himself says of the Feynman propagator $D_F$ in eqn (7): "The real part [$\bar{D}$] gives rise to the self-energy and level shift, whilst the imaginary part [$D_1$] gives the level width, or transition rate for real photon emission..." (Davies 1972, p. 1027). Here, Davies uses the customary term "self-energy," but in fact, no real energy is conveyed by the time-symmetric propagator; it only conveys force, so the term "self-action" is more accurate.

electromagnetic field $A_\mu$ to zero.

Again, this does *not* mean that real photons are disallowed, an inference that leads to confusion in Davies' account. As emphasized above, in the QDAT, the only way one obtains a real photon at all is through *both* emission and absorption, i.e., the participation of both the emitter and absorber(s) in the NU-interaction. The creation of the real photon field can be quantified in terms of a complete set of field components propagating between the emitter and absorber(s); this has been presented in Kastner and Cramer (2018) and is reviewed below. In effect, the generation of a complete set of emitter/absorber fields with an appropriate phase relationship is the entire content of the quantum LTB condition.

Davies views the existence of the $D_1$ term in the context of the restriction (16) as paradoxical, since he identifies the term "real photon" solely with an external photon.[14] If we let go of that restriction (as was justified above in our observation that a real photon can indeed be emitted and absorbed), we find that real photons are indeed transferred between currents via the $D_1$ term. In fact, Davies (1972) gives a quantitative account of how this occurs (although he hesitates to acknowledge those "internal" photons as real photons, calling the relevant construction "formal"). We now review that account.

First, Davies notes the property

$$D^+(x-y) = -D^-(y-x) \tag{18}$$

which is useful in what follows. Looking again at the expression from (13) for the first-order interaction,

$$\frac{1}{2}\sum_{i,j}\int j_i^\mu(x)D_F(x-y)j_{\mu,j}(y)\,d^4x\,d^4y \tag{19}$$

This is the first-order term in the *S* matrix, corresponding to the exchange of one photon (either virtual or real, since $D_F$ does not make this distinction). Using the decomposition (15) for $D_F$, we have:

$$\frac{1}{2}\sum_{i,j}\int j_i^\mu(x)\left(\bar{D}(x-y) - \frac{i}{2}D_1(x-y)\right)j_{\mu,j}(y)\,d^4x\,d^4y \tag{20}$$

---

[14] Davies (1972, p. 1027) suggests that real photons can interfere with virtual photons, resulting in cancellation of the advanced effects of a real photon (which he assumes has an infinite lifetime). But this is only a semi-classical argument that does not carry over into the fully quantum form of the DAT, since different photons do not mutually interfere; and certainly not photons with different physical status regarding whether or not they are on the mass shell. This is also evident from the form of (17), in which different external photon states must vanish separately. Davies appeals to a semi-classical argument because he doesn't acknowledge that one can have a real, but 'internal,' photon.

As Davies notes, the first term (real part) gives us the basic time-symmetric interaction corresponding to off-shell (virtual) photons, while the second term (imaginary part) corresponds to on-shell, real photons. The imaginary part can be written in terms of (9) as:

$$\frac{1}{4}\sum_{i,j}\int j_i^\mu(x)\left(D^+(x-y)-D^-(x-y)\right)j_{\mu,j}(y)\, d^4x\, d^4y, \qquad (21)$$

which using property (18) becomes

$$\frac{1}{4}\sum_{i,j}\int j_i^\mu(x)\left(D^+(x-y)+D^+(y-x)\right)j_{\mu,j}(y)\, d^4x\, d^4y. \qquad (22)$$

Because of the double summation over $i, j$, the two terms are the same, so we are left with:

$$\frac{1}{2}\sum_{i,j}\int j_i^\mu(x)D^+(x-y)j_{\mu,j}(y)\, d^4x\, d^4y \qquad (23)$$

In other words, the Feynman propagator leads to the radiation of positive frequencies only. (The opposite phase relationship between the fields generated by emitters and absorbers would lead to the Dyson propagator, with negative frequencies being radiated.)

Now, the final step is to note that $D^+$ in the integrand of (23) factorizes into a sum over momenta, i.e.:

$$-D^+(x-y) = i\langle 0|\hat{A}(x)\hat{A}(y)|0\rangle = i\sum_k \langle 0|\hat{A}(x)|k\rangle\langle k|\hat{A}(y)|0\rangle \qquad (24)$$

Again, this represents a real photon, since the action of each of the creation and annihilation operators in $A$ is to create and to annihilate a real, on-shell photon in mode $k$. But the photon can only end up going to one absorber, not to the many different absorbers implied by the sum, so this is why there has to be "collapse" or reduction, with an attendant probability for each possible outcome. Again, we only get this factorizable "free field" in the presence of the NU-interaction (or what is called "absorber response" at the non-relativistic level). Thus, quantization arises not from a pre-existing system of oscillators, but rather from a specific kind of field interaction--i.e., the NU-interaction. Note that the right-hand side of (24) describes a sum over products of conjugate transition amplitudes for states of varying momenta.[15] This reflects the fact that a real photon is not really a Fock state, which designates only an offer wave, but is really a squared quantity--essentially the vacuum expectation value of a projection operator. Thus, we clearly see the physical origin of the squaring procedure of the Born Rule: the photon is created through an interaction among emitter *and* absorbers (not unilaterally), but ultimately can only be delivered to one of the absorbers. In addition, this product form gives us the correct units for the photon; i.e., energy, whereas the units corresponding to a ket alone are the square root of energy.

---

[15] The two amplitudes have different spacetime arguments, so there is an overall phase factor reflecting the emission and absorption loci with respect to the inertial frame in which the fields are defined. The photon itself has no inertial frame and is described only by the conserved currents it transfers, corresponding to the square of the field amplitude. Thus the phase factor applies to the fermionic field sources. The different roles of photons and fermions in RTI is discussed further in Kastner (2019).

In light of (24), the quantum version of the "light tight box" condition is simply the completeness condition: i.e., the fact that the factorization over quantum states of a transferred photon can only be carried out if the set of states is complete. Physically, this means that absorbers corresponding to each possible value of *k* must respond; or, more accurately at the relativistic level, that the emitter and absorbers must engage in a mutual interaction, above and beyond the off-shell time-symmetric field $\bar{D}$, to generate an on-shell field that can be factorized, corresponding to the quantum completeness condition.

There is a bit of a subtlety here in understanding what counts as a "complete set" of momenta. Typically, one assumes a continuum of momentum values, but this is a mathematical idealization that does not apply to physically realistic situations, and in particular not to the QDAT. All that is required is that all momentum projectors $|k_i\rangle\langle k_i|$ for the fields exchanged between the emitter and absorbers *i={1,N}* sum to the identity. A particular $k_j$ refers to a particular absorber *j* that engages with the emitter to jointly create one component of the on-shell field whose quantum state can be written as $|\Psi\rangle = \sum_i \langle k_i|\Psi\rangle|k_i\rangle$. Thus, these states $|k_i\rangle$ have finite spread corresponding to the effective cross-section of each absorber and any uncertainty in the photon energy.

Even though all *N* absorbers contribute to create the on-shell field, as noted above, the real photon can ultimately be received by only one absorber, and this corresponds to non-unitary state reduction to the value $k_j$ for the received photon, with the probability $|\langle k_j|\Psi\rangle|^2$ (the Born Rule). Thus, besides the elimination of the independent system of field oscillators represented by the quantized field $\hat{A}$, the entire content of the quantum LTB is just the quantum completeness condition and the phase relationship that selects the Feynman rather than Dyson propagator.[16]

---

[16] The two choices of phasing of absorber response reflect the fact that the theory has two semi-groups. These are actually empirically indistinguishable. For the Feynman propagator, bound states are built on positive energies; for the Dyson propagator, bound states are built on negative energies. Thus, any observer would see an arrow of time/energy pointing to what they would consider 'the future,' and what constitutes 'positive' or 'negative' energy is only a convention based on the structure of the bound states. Here we differ with Davies (1972, pp. 1022-4), who suggests that the two choices are not the time-inverse of one another. That conclusion follows only if one retains the positive-energy structure of bound states while employing the Dyson propagator. But arguably, that is not appropriate.

# 6 Non-unitarity

The S-matrix is unitary if all interacting currents are included in the sum (13) such that all state transitions involving those currents start from the photon vacuum state and return to the photon vacuum state. In this case, the net "free field" vanishes because of the QDAT condition disallowing truly unsourced photon states (16). However, for a subset of interacting currents, the S-matrix contains a non-unitary component: that of the "free field" $D_1$. While Davies (1972) found this feature "puzzling," the present author has noted that this element of non-unitarity provides a natural account of the measurement transition (Kastner 2015), Kastner and Cramer (2018).

The non-unitary property of the S-matrix in a vacuum-to-vacuum transition for a subset of all interacting currents is also discussed by Breuer and Petruccioni (2000), pp. 40-41. In a study of decoherence, these authors take note of the fact that the Feynman propagator is complex and contains an imaginary component of the action based on the effective 'free field' $D_1$. For a single current, the vacuum-to-vacuum scattering amplitude $S(D_1)$ corresponding to this component is:

$$S(D_1) = \exp\left(-\frac{1}{4}\int j^\mu(x) D_1(x-y) j_\mu(y) d^4x d^4y\right) \quad (25)$$

The integral in the exponential is real and positive, and can be interpreted as half the average number of photons $\bar{n}$ emitted by the current (and absorbed by another current). , The vacuum-to-vacuum probability associated with the free photon field is

$$|S(D_1)|^2 = e^{-\bar{n}} < 1 \quad (26)$$

which corresponds to the probability that no photon is emitted by the current (note that if $\bar{n} = 0$, the probability is unity). Note that this is an explicit violation of unitarity at the level of the S-matrix for a single current (i.e., when final absorption of the emitted photon(s) by other current(s) is not taken into account). Based on this result, Breuer and Petruccioni note that it is the $D_1$ component that leads to decoherence. The present author discusses the crucial dependence of decoherence on non-unitarity in Kastner (2020).

Davies further notes that the complement of (26) is the probability of photon emission by the current:

$$1 - |S(D_1)|^2 = 1 - e^{-\bar{n}} = \sum_{m=1}^{\infty} e^{-\bar{n}} \frac{\bar{n}^m}{m!}, \quad (27)$$

where each term in the sum is the probability of emission of $m$ photon(s), the Poisson distribution applicable to the well-known infrared divergence.

# 7 Conclusion

This paper has discussed the fully quantum version of the direct-action (absorber) theory of fields originally proposed by Wheeler and Feynman. It has been argued that the so-called "light tight box" (LTB) condition applying to the direct-action theory needs critical review even at the classical level, and requires explicit revision at the quantum level. The condition at the classical level, (4), can be interpreted to mean that there is no truly unsourced radiation field, rather than the usual interpretation that "all emitted radiation is absorbed," since the condition actually says nothing about absorption, but says only that the net free field is zero. At the quantum level (QDAT), the boundary condition is correctly represented by (17), which simply says that there exist no true "external" photon states. A theorem showing the equivalence between the standard quantized field theory and the QDAT reveals that the condition is simply the vanishing of the quantized field $\hat{A}_\mu$. Instead, in the QDAT, interactions are mediated by a non-quantized electromagnetic potential that directly connects charged currents through the time-symmetric propagator. Quantization then arises through the relativistic analog of "absorber response," a non-unitary process whose occurrence is governed by a well-defined probability, proportional to the fine structure constant. Such a process corresponds to the generation of a real, on-shell photon.

In order to understand the conditions for real photon generation in the QDAT, it must be understood that a real, on-shell photon can indeed be emitted and absorbed and therefore be "internal," with a finite lifetime. Under a form of the quantum completeness condition, and governed by the fine-structure constant (and relevant transition probabilities), an effective "free field" corresponding to the even homogeneous solution, $D_1$, can arise. This is the quantum analog of "absorber response," which at the relativistic level is a mutual non-unitary interaction between emitter and absorber(s) that gives rise to one or more real, on-shell photons, even though such photons are technically "internal." The presence of $D_1$ converts the time-symmetric propagator into the usual Feynman propagator (eqn. 11). No "light tight box," i.e., no particular configuration of absorbers, is required for these processes to occur, so that no particular cosmological conditions need obtain in order for the QDAT to be fully applicable.

Finally, it should be remarked that historically, the direct-action theory has been subject to something of a stigma based on the fact that its primary developers, Wheeler and Feynman, abandoned their theory after they found that some form of self-action was required. However, there is nothing technically wrong with the theory, and (besides its utility in resolving consistency problems in standard quantum field theory, cf. Kastner, 2015) it should be noted that in 2003, Wheeler had returned to it as a way forward for solving the quantum gravity problem. In a joint work with Wesley, he noted that:

> [The Wheeler-Feynman theory] swept the electromagnetic field from between the charged
> particles and replaced it with "half-retarded, half advanced direct interaction" between particle
> and particle. It was the high point of this work to show that the standard and well-tested force of reaction of radiation on an accelerated charge is accounted for as the sum of the direct actions on that charge by all the charges of any distant complete absorber. Such a formulation enforces global physical laws, and results in a quantitatively correct description of radiative phenomena, without assigning stress-energy to the electromagnetic field. (Wesley and Wheeler 2003, p. 427)